\documentstyle[sprocl]{article}

\def\NP{{\em Nucl. Phys.}}
\def\NPA{{\em Nucl. Phys.} A}
\def\NPB{{\em Nucl. Phys.} B}
\def\PL{{\em Phys. Lett.}}
\def\PLB{{\em Phys. Lett.}  B}
\def\PRL{{\em Phys. Rev. Lett.}}
\def\PR{{\em Phys. Rev.}}
\def\PRC{{\em Phys. Rev.} C}
\def\PRD{{\em Phys. Rev.} D}

\def\ra{\rightarrow}

\def\be{\begin{equation}}
\def\ee{\end{equation}}
\def\bea{\begin{eqnarray}}
\def\eea{\end{eqnarray}}

\def\bfpi{\mbox{\boldmath $\pi$}}

\def\bfsigma{\mbox{\boldmath $\sigma$}}
\def\bftau{\mbox{\boldmath $\tau$}}

\def\noin{\noindent}
\def\non{\nonumber}
\def\lab{\label}
\def\bib{\bibitem}

\def\be{\begin{equation}}
\def\ee{\end{equation}}
\def\bea {\begin{eqnarray}}
\def\eea {\end{eqnarray}}

\def\dag{\dagger}

\def\ra{\rightarrow}

\def\del{\partial}

\def\lsim{\mbox{{\scriptsize \raisebox{-.9ex}
      {$\;\stackrel{{\textstyle <}}{\sim}\,$} }} }

\def\vslash{\not\!\!\mbox{\large $v$}}
\def\hyphen{{\mbox{-}}}

\def\rme{{\rm e}}\def\rmi{{\rm i}}

\def\mN{m_{\mbox{\tiny N}}}

\def\fV{f_{\mbox{\tiny V}}}
\def\fA{f_{\mbox{\tiny A}}}

\def\fM{f_{\mbox{\tiny M}}}

\def\fP{f_{\mbox{\tiny P}}}
\def\gP{g_{\mbox{\tiny P}}}

\def\GRMC{\Gamma_{\mbox{\tiny RMC}}}

\def\cL{{\cal L}}\def\cM{{\cal M}}
\def\cO{{\cal O}}

\def\gam{\gamma}

\def\lam{\lambda}
\def\Lam{\Lambda}

\def\bfnabla{\mbox{\boldmath $\nabla$}}

\def\CPT{{\small $\chi$PT}}
\def\cLHB{\cL^{\mbox{\scriptsize HB}}_{\rm ch}}

\def\ie{{\it i.e.\ }}

\def\eg{{\it e.\ g.\ }}
\def\etal{{\it et al.\ }}

\begin{document}

\title{CURRENT ISSUES IN THE NUCLEAR CHIRAL PERTURBATION 
APPROACH\footnote{invited talk at the APCTP Workshop on Astro-Hadron Physics,
Seoul, Korea, October 1997}}

\author{K. KUBODERA}

\address{Department of Physics and Astronomy, 
University of South Carolina, Columbia,\\SC 29208,
USA\\E-mail: kubodera@nuc003.psc.sc.edu}

\maketitle\abstracts{After outlining the basic
ideas of nuclear chiral perturbation theory,
I discuss its application,
presenting three examples which I believe 
are of great current interest:
(1) Exchange currents in those cases where
the leading chiral-order terms dominate;
(2) The near-threshold
$p+p\ra p+p+\pi^0$ reaction,
for which the leading order terms are suppressed 
and consequently higher-order corrections play
a prominent role;
(3) Radiative $\mu$-capture on the proton.}

\section{Introduction}

Understanding nuclear dynamics 
in relation to the fundamental QCD is one of the most
important research frontiers
in contemporary nuclear physics.
In our endeavor in this direction
chiral perturbation theory (\CPT) offers
a valuable guiding principle.
In this talk,
after giving a minimalist summary of \CPT,
I describe three examples
of the nuclear physics applications of \CPT.
I first discuss the exchange currents 
in the space components of 
the electromagnetic current and
in the time component of the axial current.
Next I discuss the use of \CPT\ 
in describing the $p+p\ra p+p+\pi^0$ 
reaction near threshold.
Finally, I describe an application of \CPT\ 
to the radiative $\mu$ capture on the proton.
These examples, I hope, will illustrate
the usefulness and versatility of nuclear \CPT.

The basic idea of effective theory,\cite{gl84}
of which \CPT\  is an example,\cite{gl84,bkm95}
is not completely foreign to nuclear physicists.
For example, in order to truncate
nuclear Hilbert space down to
a manageable model space,
we often {\it tame} the highly singular $N$-$N$ interactions
by first summing up the contributions
of high excited states that lie outside 
the chosen model space.
We then use the resulting $G$-matrices
as effective interactions
acting on the model space.\cite{bh58}
The basic picture of effective theories\cite{gl84,bkm95}
is rather similar to this nuclear physics example,
and the introduction of \CPT\  follows 
a general pattern of effective theories.
Let us consider a path integral expression
for a vacuum-to-vacuum amplitude
in QCD in the presence of external fields
\be
\rme^{iZ[v,a,s,p]}=
\int[d{\mbox{\footnotesize$G$}}]
[dq] [d\bar{q}]\,
\rme^{\rmi\int \!d^4\!x\,
\cL(q,{\bar q},G;\,v,a,s,p)}
\lab{eq:ZQCDsource}
\ee
where
$\cL=\cL^0_{\rm QCD}+\bar{q}\gamma^\mu
[v_\mu(x)-\gamma^5 a_\mu(x)]q
-\bar{q}[s(x)-ip(x)]q$.
The external fields, 
$v_\mu$, $a_\mu$, $s$ and $p$,
are endowed with appropriate SU($N$)$\times$SU($N$)
transformation properties 
to make $\cL$ chiral invariant.
(In our case, $N=2$.)
Suppose we are interested in
low-energy phenomena of pions
with their typical energy scales 
$E\lsim\!\Lam_\chi\!\sim$1 GeV,
where $\Lam_\chi$ is the QCD scale.
An effective Lagrangian that describes
low-energy behavior of QCD 
involves the Goldstone bosons and 
is introduced through 
\be
\rme^{iZ[v,a,s,p]}=
\int[d\mbox{\footnotesize{$U$}}]\,
\rme^{\rmi\int \!d^4\!x\,
\cL_{\rm ch}(U;\,v,a,s,p)},
\lab{eq:ZLeff}
\ee
where 
$U\equiv \exp(i\sum_{a=1}^3\pi^a\tau^a/f_\pi)$
with $\pi^a$'s representing the pion fields.
(Another example of many possible choices of $U$ is\cite{bkm95}
$U(x) = \sqrt{1-[\bfpi(x)/f_\pi]^2}
+i\bftau \!\cdot\! \bfpi(x)/f_\pi$.)
In \CPT, $\cL_{\rm ch}$ is expanded
in powers of $\del_\mu/\Lam_\chi$ and 
the quark mass matrix $\cM/\Lam_\chi$
and, for a given order of expansion,
all terms that are consistent with 
the symmetries are retained.
Low energy phenomena may be characterized 
by a generic pion momentum $Q$, 
which is small compared to the chiral scale 
$\Lam_\chi \sim $  1 GeV.
This suggests the possibility
of describing low-energy phenomena
in terms of ${\cal L}_{\rm{ch}}$
that contains only rather limited number of terms.
This is the basic idea of $\chi$PT.
When we try to extend this scheme 
to the baryon field $N$, 
we realize that $\del_0$ 
acting on $N$ yields $\sim\mN$(= nucelon mass),
which unfortunately is
not small compared with $\Lam_\chi$.
The heavy-baryon formalism
(HBF) was invented to avoid 
this difficulty.\cite{jm91,bkm95}
In HBF, instead of the ordinary Dirac field $N$
one works with $B$ defined by
$B(x)\equiv\rme^{\rmi \mN(v\cdot x)}N(x)$
with $v\sim(1,0,0,0)$,
shifting the energy reference point from 0 to $\mN$.
Insofar as we are only concerned 
with small energy-momenta $Q$
around this new origin,
the antibaryon may be ``integrated away".
$\cL_{\rm ch}(B, U;\,v,a,s,p)$
describing this particle-only world
may be defined similarly to Eq.(\ref{eq:ZLeff}).
The equation of motion for $B$ 
resulting from $\cL_{\rm ch}(B, U;\,v,a,s,p)$
can be rewritten as coupled equations
for the large and small components $B_{\pm}$
defined by 
$B_{\pm}\equiv P_{\pm}B$ with
$P_{\pm}\equiv (1\pm\!\vslash)/2$.
Elimination of $B_-$ in favor of $B_+$
leads to an equation of motion for $B_+$.
The HBF Lagrangian $\cLHB$ 
is defined as an effective Lagrangian 
that reproduces the equation of motion 
for $B_+$ and $U$.
Since $B_{-} \approx (Q/\mN)B_+$,
$\cLHB$ involves expansion 
in $\del_\mu/\mN$
as well as in $\del_\mu/\Lam_\chi$ 
and $\cM/\Lam_\chi$.
As $\mN\approx$ 1 GeV $\approx\Lam_\chi$,
we usually lump together chiral 
and heavy-baryon expansions.
In this combined expansion scheme,
the effective chiral Lagrangian can be organized as
\be
\cLHB=\cL^{(0)}+\cL^{(1)}+\cL^{(2)}+\,\cdots\;\;\;\;
\,,\;\;\;\;\;\;
\cL^{(\bar{\nu})}=\cO(Q^{\bar{\nu}})
\lab{eq:cLHB}
\ee
The chiral order index ${\bar \nu}$ in HBF is defined as
\be
\bar{\nu}=d+(n/2)-2,
\lab{eq:nubar}
\ee
where $n$ is the number of 
fermion lines that participate in a vertex,
and $d$ is the number of derivatives
(with $\cM\propto m_\pi^2$ counted 
as two derivatives). 
The leading order terms are given as\cite{bkm95}
\bea
{\cal L}^{(0)}&= &
 \frac{f^2_\pi}{4} \mbox{Tr} 
[ \partial_\mu U^\dagger \partial^\mu U 
 + m_\pi^2 (U^\dagger +  U - 2) ] \non
\\ &&
 + \bar{B} ( i v \cdot D + g_A^{} S \cdot u ) B 
 - \frac12 \displaystyle \sum_A 
      C_A (\bar{B} \Gamma_A B)^2 
\lab{eq17a}\\
{\cal L}^{(1)} &= &
-\frac{i g_A^{}}{2\mN} \bar{B} 
\{ S \!\cdot\! D, v \!\cdot\! u \} B 
 + 2c_1 m_\pi^2 \bar{B} B \mbox{Tr} 
( U + U^\dagger - 2 ) \non
\\ &&
 + (c_2 \!-\! \frac{g_A^2}{8\mN}) \bar{B} 
(v \!\cdot\! u)^2 B 
 + c_3 \bar{B} u \!\cdot\! u  B  \non
\\ &&
  - \frac{c_9}{2\mN} (\bar{B}B)
(\bar{B} i S \!\cdot\! u B ) 
  - \frac{c_{10}}{2\mN} (\bar{B} S^\mu B) 
(\bar{B} i u_\mu B).
\lab{eq17b}
\eea
Here $\xi=\sqrt{U}$,
$u_\mu \equiv i (\xi^\dag \del_\mu \xi
                - \xi \del_\mu \xi^\dag)$, 
$S_\mu=i\gam_5\sigma_{\mu\nu}v^\nu/2$ and 
\be
D_\mu=\del_\mu+[\xi^\dag,\del_\mu u]/2
-i\,\xi^\dag(v_\mu+a_\mu)\xi/2
-i\xi(v_\mu-a_\mu)\xi^\dag/2.
\ee
The four-velocity parameter $v_\mu$
is in practice chosen to be $v_\mu=(1,0,0,0)$.
The low-energy constants $c_1,c_2$ and $c_3$
have been determined 
from phenomenology,\cite{bkm95,pmmmk96} 
and their numerical values
will be discussed later.\footnote{The most updated determination 
of $\cLHB $ can be found in Ref.\cite{eckmoj}.}
In any practical calculations, 
one expands $U(x)$ in powers of
$\bfpi(x)/f_\pi$ and
and only retains necessary lowest order terms.

The chiral counting above 
applies to individual vertices.
We can also introduce chiral counting 
for Feynman diagrams;
the chiral order $\nu$ of a Feynman diagram 
is given by\cite{wei90}
\be
\nu\,=\,2-N_E+2L-2(C-1)+\sum_i\bar{\nu}_i,
\lab{eq:Dcount}
\ee
where $N_E$ is the number of external fermion lines, 
$L$ the number of loops, 
$C$ the number of disconnected parts,
and the sum runs over vertices
involved in the Feynman diagram.
In fact, in applying this counting rule
to nuclear physics,
we must exercise caution.
As emphasized by Weinberg,\cite{wei90}
naive chiral counting fails for a nucleus, 
because purely nucleonic intermediate states
(with no pions in flight) that occur in a nucleus
can have very low excitation energies
and thereby invalidate the ordinary chiral counting rule.
To avoid this difficulty,
we need to classify Feynman diagrams into two groups.
Diagrams in which every intermediate state
contains at least one meson in flight
are called irreducible diagrams,
and all others are called reducible diagrams.
The \CPT\  can only be applied 
to the irreducible diagrams.
One can show that Eq.(\ref{eq:Dcount}) 
gives a correct chiral order for an irreducible diagram,
and that an irreducible diagram
of chiral order $\nu$ is typically $\cO(Q^{\nu})$.
The contribution of 
all the irreducible diagrams 
(up to a specified chiral order)
is then used as an effective operator
acting on the nucleonic Hilbert space.
This second step allows us to incorporates 
the contributions of the reducible diagrams
to infinite order.
This two-step procedure may be referred to as
{\it nuclear chiral perturbation theory}.

Weinberg\cite{wei90} formulated nuclear \CPT\ 
and applied it to a chiral perturbative derivation 
of the nucleon-nucleon interactions.
Further developments by van Kolck \etal\cite{kol92}
succeeded in explaining many basic features of
the two-nucleon interactions,
but their phenomenological success 
has not quite reached the level of
the traditional phenomenological boson-exchange 
potentials.

Nuclear \CPT\  can also be applied to 
electroweak transition processes.\cite{rho91}
Here a nuclear transition operator is defined 
as a set of all the irreducible diagrams
(up to a given chiral order $\nu$) 
with an external current inserted.
In a completely consistent \CPT\ calculation, 
this transition operator
is to be sandwiched between the initial and final 
nuclear states which are governed 
by the nucleon interactions corresponding to
the $\nu$-th order irreducible diagrams.  
However, the present status of the 
\CPT\  derivation of the $N$-$N$ forces
is such that we cannot completely fulfill
this formal consistency.
In practice, therefore, we use the phenomenological 
nucleon-nucleon interactions 
in the nuclear Schr\"odinger equation
to generate the initial and final nuclear wavefunctions.
We shall refer to this eclectic method
as the {\it hybrid treatment} of nuclear \CPT.

\section{Exchange currents}

First, I wish to talk about \CPT treatments of 
exchange currents for those pieces
of the electroweak currents 
for which the one-pion exchange contributions
are known to be dominant. 
As discussed in Ref.\cite{kdr78},
the space component of the vector current (${\bf V}$) and
the time component of the axial current ($\!A_0\!$)
belong to this category.
From the \CPT\  point of view,
the pion-exchange current that arises
from the low-energy theorem\cite{cr71} 
corresponds to the leading-order tree diagram 
in chiral expansion
applied to a two-nucleon system.\cite{rho91}
Furthermore, \CPT\  provides a systematic framework
for organizing exchange-current contributions
that have shorter ranges than the one-pion range.\cite{rho91,pmr93,pmr95}

As regards ${\bf V}$, there is a very clean test
in the two-nucleon systems,
for which the nuclear wavefunctions
are known with sufficient accuracies.
Experimentally, the capture rate for the 
$n({\rm thermal})+p \ra d+\gamma$ process is
$\sigma_{\rm exp}=334.2\!\pm\!0.5$ mb,
and this value is $\sim$10\% larger than the 
impulse approximation (IA) prediction,
$\sigma_{IA}=302.5\!\pm\!4.0$mb. 
Riska and Brown\cite{rb72} pointed out
that the one-pion exchange current derived
from the low-energy theorem\cite{cr71}
can account for $\sim$70\% of the 
missing capture rate.
Another impressive success of 
the exchange current calculations
based on the low-energy theorem
is known for the $e+d\ra e+p+n$ reaction,
see \eg \,Ref.\cite{fm89}.
Park \etal,\cite{pmr95} carried out
a systematic study of the contributions of 
the next-to-leading order terms in chiral expansion.
According to this study, the inclusion of the next order
contributions leads to $\sigma=334\!\pm\!2$ mb
in perfect agreement with experiment.

Regarding $A_0$ 
(time component of the axial current),
it is not easy to find
observables in the two-nucleon system
that give clear-cut information 
about  the leading one-pion exchange current.
This is mainly because $A_0$ must compete 
with the dominant space component ${\bf A}$.
We therefore need to go to complex nuclei.
This requires careful examinations of
the so-called core polarization effects.
Despite this non-trivial aspect, 
there is by now ample evidence 
that supports the \CPT\  derivation of exchange currents.
Warburton \etal's 
systematic analyses\cite{war91,wtb94}
of the first-forbidden $\beta$ transitions indicate that, 
over a very wide range of the periodic table,
the ratio of the exchange-current contribution
to the 1-body contribution 
$\delta_{\rm{mec}}\equiv
\langle A^0({\rm mec})\rangle/
\langle\!A^0(1\hyphen{\rm body})\rangle$
is 
$\delta_{\rm{mec}}^{\rm exp}\!=\!0.5 \sim 0.8$.
The leading-order \CPT\  term,
\ie the soft-pion exchange term\cite{kdr78,rho91}
can explain the bulk of 
$\delta_{\rm{mec}}^{\rm exp}$
and the next-order \CPT\  term\cite{pmr93,ptk94}
gives an additional $\sim$10\% enhancement,
bringing the theoretical value closer to 
the empirical value.
The general tendency that
$\delta_{\rm{mec}}^{\rm theor}$ is 
still somewhat smaller than 
$\delta_{\rm{mec}}^{\rm exp}$
has been a subject of recent intensive studies.
I wish to mention, however,
that in the A=16 nuclei,\cite{wtb94}
where the shell model space employed is large enough
to get rid of the usual core-polarization corrections
and therefore the calculation is deemed most reliable,
the lowest-order one-pion current gives 
good agreement with the data.
Thus, the ``extra enhancement" of 
$\delta_{\rm{mec}}$ for the heavier nuclei
should be taken with some caution.

If we take the putative extra enhancement
in $\delta_{\rm{mec}}$ seriously,
what is a possible mechanism for that ?
Several authors investigated heavy-meson 
exchange contributions.\cite{krt92}
In particular, Towner\cite{krt92} carried out
a detailed calculation using the ``hard-pion formalism"
in conjunction with the Lagrangian 
that engenders the phenemenological 
$N$-$N$ interactions,
and was able to obtain a roughly right amount of
extra enhancement to reproduce 
$\delta_{\rm{mec}}^{\rm exp}$.
Meanwhile, the use of a in-medium chiral Lagrangian 
that by fiat incorporates the BR scaling\cite{br91}
was also proposed as a possible solution
to the extra enhancement problem.\cite{kr91}
These two explanations in fact may not be 
as disjoint as they appear.
The most prominent heavy-meson pair contribution
comes from $\sigma$-exchange,
whereas this $\sigma$ contribution 
can be effectively rewritten 
as the $1$-body term with the nucleon mass 
replaced by an effective mass.\cite{dt87}
Thus the $\sigma$-meson contribution
plays a role similar to the BR scaling.\cite{br91}
So we do have at our disposal some models
that can accommodate a certain amount of 
the ``extra enhancement".
What is not clear is their relation
to the basic chiral counting.
For example, if we are to interpret the $\sigma$-meson 
as multipion-exchange effects, 
they would correspond 
to much elevated chiral orders, 
and there is at present no consistent way to treat
all possible terms of such high chiral orders.
Meanwhile, some multi-fermion terms
that have higher chiral orders than
the next-to-leading-order terms
give rise to nucleon mass shift in medium.
In this sense the BR scaling has some overlap 
with higher-order terms in \CPT, 
but again its systematic treatment in \CPT\  
is at present not available.

Apart from these possible higher chiral order effects,
I wish to mention a subtle point that exists even 
within the next-to-leading-order 
calculations which scored 
great success.\cite{pmr93,pmr95,ptk94}
If we follow faithfully the formal chiral counting rules,
transition operators in general can contain 
terms of the contact multi-fermion type
as well as extremely short-ranged terms that 
reflect the integrated-out high-energy physics.
However, in the existing treatments
which are based on the {\it hybrid treatment}
of nuclear \CPT,
these singular terms are discarded 
using the intuitive argument
that the short-range $N$-$N$ correlation
would strongly suppress these zero-range or
extremely shor-ranged transition 
operators.\cite{pmr93,ptk94}

One of the most urgent problems
in nuclear \CPT\  at present is
how to justify, or how to get rid of 
this intuitive argument.
To make progress in this direction,
we must learn how to apply \CPT\  to 
short-range phenomena.
We also need to treat on the same footing
the irreducible kernels that generate 
effective electroweak transition opertaors
and the irreducible kernels 
which are responsible for distortion
of the initial and final nuclear wavefunction.
As a useful testing ground for this type of 
fully consistent nuclear \CPT calculations
(free from the {\it  hybrid treatment}),
we can consider a system 
with a characteristic energy low enough 
for even the pions to be integrated out.
In particular, it is interesting to study
whether we can describe low-energy two-nucleon
phenomena consistently in a version of nuclear \CPT\ 
in which the pionic degrees of freedom
have been integrated out and their effects
are incorporated into the low-energy coeffcients
of an effective Lagangian for the nucleons.
A \CPT\  calculation without pions
is almost an oxymoron but, 
in the formal sense of heavy-baryon \CPT,
nothing prevents us from integrating out pions
from our system.
Also, as a first step toward the eventual 
reintroduction of the pionic degrees of freedom,
this nucleon-only \CPT\  is expected to be a
very informative exercise.

As for the low-energy N-N scattering,
Kaplan, Savage and Wise\cite{ksw96}
recently attempted to extend \CPT\  
to description of the short-range $N$-$N$ interactions.
The result, however,  was not very encouraging 
in that the chiral expansion turned out to have a poor convergence property.
But it was soon realized\cite{lm97,bcp97}
that a dimensional regularization used 
in Ref.\cite{ksw96} is responsible for this difficulty.
The use of a cut-off regularization 
largely eliminates the problem.\cite{bcp97,lep97}

A challenge now is to carry out a similar calculation
for the two-nucleon observables that involve 
external probes.
Very recently, Park \etal,\cite{pkmr97}
has carried out the first study of this type.
Dr. Tae-Sun Park will give a detailed account of that work
in his talk right after mine.

\section{$p+p \ra p+p+\pi^0$ reaction near threshold}

My next topic is the low-energy 
$p+p \ra p+p+\pi^0$ reaction.
The recent high-precision measurements\cite{meyetal90}
of the total cross sections 
for $pp \rightarrow pp\pi^0$ near threshold 
have invited many theoretical investigations
on this process.
I attempt here to explain 
why this reaction is of particular interest 
from a \CPT\  point of view.
To this end, I first give a quick general survey of
the theoretical developments 
up until 1996,\footnote{Regrettably I must 
skip here many issues that have been extensively studied 
by multiple-scattering specialists.  For a recent review
on the topics omitted here, see \eg Ref.\cite{bla90}.
I am indebted to Tony Thomas and Boris Blankleider 
for bringing my attention to the line of work
described in this reference.}
and subsequently I will expound 
on our latest study.\cite{slmk97}
At the end I present some precautionary remarks 
concerning the results 
described in Ref.\cite{slmk97}.

One expects that threshold pion production occurs
via the single-nucleon process 
(the impulse or Born term), fig.1(a),
and the $s$-wave pion rescattering process, fig.1(b).


\begin{picture}(300,145)


\put(0,40){\line(1,0){20}}
\put(30,40){\line(1,0){50}}
\put(90,40){\line(1,0){20}}

\put(0,90){\line(1,0){20}}
\put(30,90){\line(1,0){50}}
\put(90,90){\line(1,0){20}}

\put(55,90){\line(1,1){10}}
\put(67,102){\line(1,1){10}}
\put(79,114){\line(1,1){10}}

\put(25,65){\oval(10,60)}
\put(85,65){\oval(10,60)}

\put(46,82){\makebox(0,0){$p_2$}}
\put(64,82){\makebox(0,0){$p_2'$}}
\put(46,32){\makebox(0,0){$p_1$}}
\put(64,32){\makebox(0,0){$p_1'$}}
\put(66,110){\makebox(0,0){$q$}}

\put(93,130){\makebox(0,0){$\pi^0$}}
\put(-6,90){\makebox(0,0){$p$}}
\put(116,90){\makebox(0,0){$p$}}
\put(-6,40){\makebox(0,0){$p$}}
\put(116,40){\makebox(0,0){$p$}}

\put(10,82){\makebox(0,0){$\bar{p}_2$}}
\put(100,82){\makebox(0,0){$\bar{p}_2'$}}
\put(10,32){\makebox(0,0){$\bar{p}_1$}}
\put(100,32){\makebox(0,0){$\bar{p}_1'$}}

\put(55,0){\makebox(0,0){fig. 1(a)}}


\put(165,40){\line(1,0){20}}
\put(195,40){\line(1,0){50}}
\put(255,40){\line(1,0){20}}

\put(165,90){\line(1,0){20}}
\put(195,90){\line(1,0){50}}
\put(255,90){\line(1,0){20}}

\put(220,90){\line(1,1){10}}
\put(232,102){\line(1,1){10}}
\put(244,114){\line(1,1){10}}

\put(220,40){\line(0,1){7}}
\put(220,50){\line(0,1){7}}
\put(220,60){\line(0,1){7}}
\put(220,70){\line(0,1){7}}
\put(220,80){\line(0,1){7}}

\put(190,65){\oval(10,60)}
\put(250,65){\oval(10,60)}

\put(211,82){\makebox(0,0){$p_2$}}
\put(229,82){\makebox(0,0){$p_2'$}}
\put(211,32){\makebox(0,0){$p_1$}}
\put(229,32){\makebox(0,0){$p_1'$}}
\put(229,110){\makebox(0,0){$q$}}

\put(258,130){\makebox(0,0){$\pi^0$}}
\put(159,90){\makebox(0,0){$p$}}
\put(281,90){\makebox(0,0){$p$}}
\put(159,40){\makebox(0,0){$p$}}
\put(281,40){\makebox(0,0){$p$}}
\put(228,62){\makebox(0,0){$\pi^0$}}

\put(175,82){\makebox(0,0){$\bar{p}_2$}}
\put(265,82){\makebox(0,0){$\bar{p}_2'$}}
\put(175,32){\makebox(0,0){$\bar{p}_1$}}
\put(265,32){\makebox(0,0){$\bar{p}_1'$}}
\put(214,62){\makebox(0,0){$k$}}

\put(220,0){\makebox(0,0){fig. 1(b)}}

\end{picture}

\vspace{0.8cm}
\noin
In the conventional treatment\cite{kr66},
the $\pi$-$N$ vertex for the impulse term is
assumed to be given by the Hamiltonian
\begin{equation}
{\cal H}_0 = \frac{g_A}{2 f_\pi} \bar{\psi}
\left( \vec{\sigma} \!\cdot\!
\vec{\nabla}
(\mbox{\boldmath$\tau$} \!\cdot\!
\mbox{\boldmath$\pi$} )
- \frac{i}{2m_{\mbox{\tiny N}}}
\{ \vec{\sigma} \!\cdot\!
\vec{\nabla},
\mbox{\boldmath$\tau$} \!\cdot\!
\dot{\mbox{\boldmath$\pi$}} \}
\right) \psi,
\label{eq:H0}
\end{equation}
where $g_A$ is the
axial coupling constant,
and $f_\pi$ = 93 MeV is the pion decay constant.
The first term gives
$p$-wave pion-nucleon coupling,
while the second term accounts
for the nucleon recoil effect.
The s-wave rescattering vertex in Fig.1(b)
is customarily described with the phenomenological
Hamiltonian\cite{kr66}
\begin{equation}
{\cal H}_{1} =
4\pi \frac{\lambda_1}{m_\pi} \bar{\psi}
\mbox{\boldmath$\pi$}\!\cdot\!
\mbox{\boldmath$\pi$} \psi
+ 4\pi \frac{\lambda_2}{m^2_\pi}
\bar{\psi} \mbox{\boldmath$\tau$}\!\cdot\!
\mbox{\boldmath$\pi$} \!\times\!
\dot{\mbox{\boldmath$\pi$}} \psi
\label{eq:H1}
\end{equation}
The coupling constants
$\lambda_1$ and $\lambda_2$ determined
from the experimental pion-nucleon scattering lengths are
$\lambda_1 \sim 0.005$ and
$\lambda_2 \sim 0.05$.
Thus,  $\lambda_1 \ll \lambda_2$,
in conformity to an expectation from current algebra.
The calculations based on
these phenomenological vertices\cite{kr66}
yield cross sections for s-wave $\pi^0$ production
that are significantly smaller,
typically by a factor of $\sim$5,
than the experimental values.\cite{meyetal90}
To many people this large discrepancy 
came as a big surprise.
Apart from this discrepancy,
it is worthwhile to emphasize 
that the near-threshold $p+p \ra p+p+\pi^0$ reaction
is an intrinsically suppressed process 
for the following reasons.
First, in Eq.(\ref{eq:H0}),
only the second term contributes
to $s$-wave pion production.
The suppression factor
$\sim m_\pi /m_{\mbox{\tiny N}}$
contained in this term drastically reduces
the contribution of the impulse term, Fig.1(a),
enhancing thereby the relative importance
of the two-body rescattering process, Fig.1(b).
On the other hand, the dominant $\lambda_2$ term
in Eq.(\ref{eq:H1})
cannot contribute to the $pp\pi^0\pi^0$ vertex
in Fig.1(b) due to its isospin structure.
Thus the two-body contribution is also hindered here.
As we will see below,
the fact that the leading terms 
in the phenomenological Lagrangian,
Eqs.(\ref{eq:H0}) and (\ref{eq:H1}),
cannot contribute to the near-threshold 
$p+p \ra p+p+\pi^0$ reaction implies that
this reaction is sensitive to higher-order terms
in chiral expansion.
This is one of the reasons why this reaction 
is particularly interesting from the \CPT\ 
point of view.

To explain major issues involved 
in the more recent theoretical developments,
it is convenient to introduce
what we call the {\it typical threshold}
({\it TT}) {\it kinematics}.
Consider Fig.1(b) in the center of mass (CM) system
with the initial and final interactions turned off.
At threshold,
$(q_0,\,\vec{q}) = (m_\pi,\,\vec{0})$,
$p_{10}'=p_{20}'=m_{\mbox{\tiny N}}$,
$\vec{p}_1^{\,\,\prime}
=\vec{p}_2^{\,\,\prime}=\vec{0}$,
so that
$p_{10}=p_{20}=m_{\mbox{\tiny N}}
+m_\pi/2$, $k_0=m_\pi/2$,
$\vec{p}_1=\vec{k}=-\vec{p}_2$ with
$|\vec{k}| = \sqrt{m_\pi m_{\mbox{\tiny N}}
+ m_\pi^2/4}$.
(Of course, even for $\vec{q} = 0$,
the actual kinematics
for the transition process may differ
from the {\it TT kinematics}
because of the initial- and final-state interactions.)
Now, for the {\it TT kinematics}, 
$k^2=-m_{\mbox{\tiny N}} m_\pi\ne m_\pi^2$,
indicating that the rescattering diagram 
is sensitive to the {\em off-shell} $\pi N$ amplitudes.
However, there is no guarantee
that ${\cal H}_1$ of Eq.(\ref{eq:H1}) describes
the off-shell amplitudes adequately.
Hern\'{a}ndez and Oset\cite{ho95}
suggested that the $s$-wave amplitude
enhanced for off-shell kinematics
could increase the rescattering contribution
sufficiently to reproduce the experimental cross sections.
However, Ref.\cite{ho95} used
phenomenological off-shell extrapolations,
the reliability of which requires further examination.
We will come back to this issue later.
Another point is that
$k^2=-m_{\mbox{\tiny N}} m_\pi$
for {\it TT kinematics}
implies that the rescattering process typically probes
inter-nucleon distances $\sim$ 0.5 fm.
The process then can be sensitive
to exchange of the heavy mesons
which play an important role
in the phenomenological meson-exchange
$N$-$N$ potentials.
Lee and Riska\cite{lr93} studied the possible
enhancement of 
the $pp \rightarrow pp\pi^0$ cross section
due to shorter-range meson exchanges.
This enhancement, however, 
turns out to be very sensitive
to how one evaluates the ``basic" diagrams,
fig. 1(a) and (b).
We will see below that a \CPT\ calculation of these diagrams
does not necessarily support 
the idea of heavy-meson exchange enhancement.

As emphasized in Introduction,
\CPT\cite{gl84,bkm95}
serves as a consistent framework
to describe the low-energy $\pi$N scattering amplitudes
for off-shell as well as on-shell kinematics.
Park \etal\cite{pmmmk96}
(to be referred to as PM$^3$K)
and Cohen \etal\cite{cfmv96}
(to be referred to as CFMK)
carried out the first $\chi$PT calculations for
the $pp \rightarrow pp\pi^0$ reaction.
The results of these two groups
essentially agree with each other
on the following major points:
(1) the pion rescattering term in a $\chi$PT treatment
is significantly larger than in the conventional treatment;
(2) the sign of the rescattering term in a $\chi$PT treatment
is opposite to that obtained in the conventional approach;
(3) the enhanced rescattering term in a $\chi$PT treatment
almost cancels the impulse term,
leading to theoretical cross sections much smaller
than the observed values;
(4) the smallness of the amplitude 
corresponding to Fig. 1(a)+(b)
implies that the addition of 
the heavy-meson exchange diagrams
does not result in large enough interference to reproduce
the experimental cross sections.
Thus \CPT\ treatments
of the off-shell $\pi N$ scattering amplitudes
indeed drastically change
the near-threshold $pp \rightarrow pp\pi^0$ cross section.

We need to mention, however,
that the calculations in Refs.\cite{pmmmk96,cfmv96},
which rely on coordinate space representation,
involve potentially problematic approximations
on the kinematical variables appearing in
the $\pi N$ scattering amplitudes.
Namely, there the r-space representation
of the two-body transition operator [fig. 1(b)]
was derived from the Feynman amplitude
corresponding to the {\it TT kinematics}
by Fourier-transforming this particular amplitude
with respect to
$\vec{p}_1$, $\vec{p}_2$, $\vec{p}^{\,\,\prime}_1$ and
$\vec{p}^{\,\,\prime}_2$, 
while keeping all the other kinematical
variables fixed at their {\it TT kinematics} values.
Although this type of kinematical simplification
is commonly used in nuclear physics
for deriving effective r-space operators,
it is expected to be much less reliable for the threshold
$pp \rightarrow pp\pi^0$ reaction.  
The reason is two-fold:
First, the energy-momentum exchange 
due to the initial and final-state interactions 
is essentially important for this process.
Secondly, the destructive interference between
the one-body and two-body terms found
in Refs.\cite{pmmmk96,cfmv96} implies that
even a rather moderate change 
in the two-body term can influence
the cross sections significantly.
In view of these problems,
Sato \etal\ \cite{slmk97} (to be referred to as SLMK)
have recently carried out
a $\chi$PT calculation for $pp \rightarrow pp\pi^0$
in the momentum representation,
which allowed them to avoid
the above-mentioned kinematical simplifications.
This improvement is found to affect strongly 
the calculated cross section.
We give below a brief summary of SLMK's work.
The necessity to discuss the basic ingredients of SLMK
also offers an opportunity to explain the earlier 
\CPT\  treatments\cite{pmmmk96,cfmv96}
in more detail than the above itemized summary allows.

In heavy-baryon \CPT,
in order to generate the one-body and
two-body diagrams depicted in figs.1(a), 1(b),
we minimally need terms with 
$\bar{\nu}=1$ and $2$ in $\cLHB$, Eq.(\ref{eq:cLHB}).
The relevant pion-nucleon interaction Hamiltonian is 
${\cal H}_{int} = {\cal H}^{(0)} + {\cal H}^{(1)}$ 
with
\bea
 {\cal H}^{(0)}&=&\frac{g_A}{2f_\pi} \bar{B} 
[ \bfsigma\!\cdot\!\bfnabla 
 ( \bftau\!\cdot\!\bfpi ) ] B 
 + \frac{1}{4f_\pi^2}\bar{B} \bftau\!\cdot\!\bfpi
\!\times\!\dot{\bfpi} B
\lab{eq:Hint0}\\
{\cal H}^{(1)}&=&
 \frac{-i g_A}{4\mN f_\pi} \bar{B} 
\{ \bfsigma\!\cdot\!\bfnabla, 
\bftau\!\cdot\!\dot{\bfpi} \} B  
\non\\  
& &+ \frac{1}{f^2_\pi} [ 2c_1 m_\pi^2 \pi^2 
 \!-\! (c_2 \!-\! \frac{g^2_A}{8\mN}) \dot{\pi}^2
 \!-\! c_3 (\partial \pi)^2 ] \bar{B} B.
\lab{eq:Hint1}
\eea
Here ${\cal H}^{(\bar{\nu})}$
represents the term of chiral order $\bar{\nu}$.
The {\it standard} values of 
the low-energy coefficients (LEC),
$c_1$, $c_2$ and $c_3$, 
may be taken from Ref.\cite{bkm95},
where these parameters
were determined from the experimental values of
the pion-nucleon $\sigma$ term,
the nucleon axial polarizability $\alpha_A$
and the isospin-even s-wave
$\pi N$ scattering length $a^+$.
Their values are 
\begin{equation}
c_1=-0.87\pm 0.11\;{\rm GeV}^{-1},\;\;
c_2=3.34\pm 0.27\;{\rm GeV}^{-1},\;\;
c_3=-5.25\pm 0.22\;{\rm GeV}^{-1}.
\label{eq:lecoef}
\end{equation}
Comparing Eqs. (\ref{eq:Hint0}), (\ref{eq:Hint1})
with the phenomenological effective Hamiltonian
${\cal H}_0+{\cal H}_1$,
Eqs.(\ref{eq:H0}),(\ref{eq:H1}), 
we note that the first term in ${\cal H}^{(0)}$
combined with the first term in ${\cal H}^{(1)}$
exactly reproduces ${\cal H}_0$.
As for the $\pi\pi\!N\!N$ vertices,
we can associate the second term in $ {\cal H}^{(0)}$
to the $\lam_2$ term in ${\cal H}_1$,
and second term in $ {\cal H}^{(1)}$
to the $\lam_1$ term in ${\cal H}_1$.
Thus, as mentioned earlier, 
the threshold $pp \rightarrow pp\pi^0$ reaction
is indeed sensitive to the higher chiral-order terms.
Concentrating on the $\lambda_1$ term,
which is of direct relevance here, 
we recognize the correspondence:
$4\pi \lam_1/m_\pi 
\Longleftrightarrow \kappa(k,q)$,
where
\be
\kappa(k,q)\equiv
\frac{m_\pi^2}{f_\pi^2} 
[2c_1 - (c_2\!-\!\frac{g_A^2}{8\mN})
\frac{\omega_q \omega_k}{m_\pi^2} 
 - c_3 \frac{q\cdot k}{m_\pi^2} ],
\lab{eq:kappakq}
\ee
with $q\!=\!(\omega_q, {\bf q})$ and 
$k\!=\!(\omega_k, {\bf k})$, see fig.1(b).
Now, for {\it on-shell} low energy pion-nucleon scattering,
{\it i.e.\/}, $k \!\sim\! q \!\sim\! (m_\pi, {\bf 0})$, 
we equate
\be
\left(
4\pi\lam_1/m_\pi
\right)_{\mbox{\scriptsize on-shell}}
^{\mbox{\tiny $\chi$PT}}
= \kappa_0\,\equiv\,
\kappa(k\!=\!(m_\pi, {\bf 0}),
q\!=\!(m_\pi, {\bf 0})),
\lab{eq:onshell}
\ee
\be
\kappa_0=\frac{m_\pi^2}{f_\pi^2}
\left(2c_1\!-\!c_2\!-\!c_3
\!+\!\frac{g_A^2}{8\mN} \right)
=-2\pi\!
\left( 1\!+\!\frac{m_\pi}{\mN} \right)\!a^+ +\!
\frac{3g_A^2}{128\pi}
\frac{m_\pi^3}{f_\pi^4}.\lab{eq:kappa1}
\ee
which gives $\kappa_0=(0.87\pm0.20)\,{\rm GeV}^{-1}$.
On the other hand, in the conventional approach,
$\lam_1$ is determined from the $a^+$ term 
in Eq.(\ref{eq:kappa1}) alone.  Namely
\be
\left(
\frac{4\pi\lam_1}{m_\pi}
\right)_{\mbox{\scriptsize conventional}}
\,=\,-2\pi\left( 1+\frac{m_\pi}{\mN} 
\right)a^+\,=\,(0.43\pm0.20)\,{\rm GeV}^{-1},
\lab{lambdaold}
\ee
which corresponds to the ``literature value"
$\lam_1\,=\,0.005$.
Thus the $\chi$PT value
$\kappa_0=0.87\,{\rm GeV}^{-1}$
is about twice as large as the conventional value. 
To pursue further comparison 
between the traditional and $\chi$PT approaches, 
let us go back to Eq.(\ref{eq:kappakq}).
Obviously, $\kappa(k,q)$
cannot be fully simulated by a constant $\lam_1$.
To illustrate how the $q$ and $k$ dependence 
in $\kappa(k,q)$ affects the rescattering vertex [fig.1(b)], 
let us consider the value of $\kappa(k,q)$
corresponding to the {\it TT kinematics}
and denote it by $\kappa_{\mbox{\tiny{TT}}}$.  
\be
\kappa_{\mbox{\tiny{TT}}}
=\frac{m_\pi^2}{f_\pi^2}\!
\left[ 2c_1\!-\!\frac{1}{2}\!
\left(\!c_2\!-\!\frac{g^2_A}{8\mN}\!\right)\!
-\!\frac{c_3}{2}
\right]
= -1.5\,{\rm GeV}^{-1}
\lab{eq:kappath}
\ee
We can see the $s$-wave $\pi$-$N$ interaction 
is much stronger here than in the on-shell cases
[Eqs.(\ref{eq:kappa1}),(\ref{lambdaold})],
and the sign of $\kappa_{\mbox{\tiny{TT}}}$
is {\it opposite} to that of $\kappa_0$.
This flip in sign changes drastically 
the interference pattern of the Born 
and rescattering terms.

Adopting nuclear $\chi$PT,
we write the transition amplitude
for the $pp \rightarrow pp\pi^0$ reaction as
\begin{equation}
T\,=\,\langle \Phi_f | {\cal T} | \Phi_i \rangle,
\label{eq:Tmatrix}
\end{equation}
where $|\Phi_i\rangle$ ($|\Phi_f\rangle$)
is the initial (final) two-nucleon state
distorted by the initial-state (final-state) interaction.
For formal consistency,
if ${\cal T}$ is calculated up to order $\nu,$
the nucleon-nucleon interactions that generate
$|\Phi_i\rangle$ and $|\Phi_f\rangle$
should also be calculated by summing up
all irreducible two-nucleon scattering diagrams
up to order $\nu$.
In practice, however, it is common
to use the phenomenological $N$-$N$
interactions that reproduce measured
two-nucleon observables,
and here again we use this hybrid version of nuclear $\chi$PT.
The lowest-order contributions
to the impulse and rescattering terms
come from the $\nu=-1$ and $\nu=1$ terms, 
respectively.
(Here we are using the counting rule 
Eq. (\ref{eq:Dcount}).)
Therefore we have
\be
{\cal T}\,=\, {\cal T}^{(-1)}+{\cal T}^{(+1)}
\equiv {\cal T}^{{\rm Imp}}
+{\cal T}^{{\rm Resc}}
\label{eq:calTtrunc}
\ee
The use of the Hamiltonian
in Eqs. (\ref{eq:Hint0}) and (\ref{eq:Hint1})
leads to momentum-space matrix elements
\begin{eqnarray}
{\cal T}^{(-1)}
& = &\frac{i}{(2\pi)^{3/2}}
\frac{1}{\sqrt{2\omega_q}}\frac{g_A}{2f_\pi}
\sum_{j=1,2} [-\vec{\sigma}_j\cdot\vec{q}
+\frac{\omega_q}{2m_N}\vec{\sigma}_j
\cdot (\vec{p}_j + \vec{p}_j^{\,\,\prime} )]\tau_j^0,
\label{eq:Tminus1} \\
{\cal T}^{(+1)}
&=& \frac{-i}{(2\pi)^{9/2}}
\frac{1}{\sqrt{2\omega_q}}\frac{g_A}{f_\pi}
\sum_{j=1,2} \kappa(k_j,q)\frac{\vec{\sigma}_j
\cdot \vec{k}_j\tau_j^0}{k_j^2 - m_\pi^2},
\label{eq:Tplus1}
\end{eqnarray}
where $\vec{p}_j$ and $\vec{p}^{\,\,\prime}_j$ ($j=1,2$)
denote the initial and final
momenta of the $j$-th proton (see fig. 1).
The four-momentum of the exchanged pion 
is defined by the nucleon four-momenta 
at the $\pi NN$ vertex:
$k_j \equiv p_j-p^{\,\,\prime}_j$, 
where 
$p_j =(E_{p_j}, \vec{p}_j),
p^{\,\,\prime}_j=(E_{p^{\,\,\prime}_j}, 
\vec{p}^{\,\,\prime}_j)$ 
with  the definition 
$E_p=(\vec{p}^{\,\,2} + m_N^2)^{1/2}$.

The transition amplitude of the
$ pp \rightarrow pp\pi^0$ reaction
is evaluated by taking the matrix element
of the production operator ${\cal T}$ defined above
between the initial ($\chi^{(+)}$)
and the final ($\chi^{(-)}$)
$pp$ scattering wavefunctions.
In terms of this transition matrix element,
the total cross section is given by
\begin{eqnarray}
\sigma_{pp\rightarrow pp\pi^0}(W)
 & = & \frac{(2\pi)^4}{2v_i}\int d\vec{p}_f d\vec{q}
       \;\delta(\sqrt{4E^2_{\vec{p}_f} + \vec{q}^2} +
\omega_q - W)  \nonumber \\
  & \times &
  \frac{1}{4}\sum_{m_{s_1}m_{s_2}m_{s'_1}m_{s'_2}}
  | <\chi^{(-)}_{\vec{p}_f, m_{s'_1},m_{s'_2}},
\vec{q}\; | {\cal T} |
      \chi^{(+)}_{\vec{p}_i, m_{s_1},m_{s_2}}> |^2,
\label{eq:sigmaWa}
\end{eqnarray}
where $\vec{p}_i$ and $\vec{p}_f$ are
the {\it asymptotic} relative momenta
of the initial and final $pp$ states, respectively,
$W=2E_{\vec{p}_i} $ is the total energy,
$v_i= 2|\vec{p}_i|/E_{\vec{p}_i}$ is the asymptotic
relative velocity of the two initial protons,
and $m_{s_j}$ is the z-component of
the spin of the $j$th nucleon.
Near threshold the transition matrix 
appearing in Eq.(\ref{eq:sigmaWa})
can be expressed using only 
one partial wave amplitude.
The corresponding reduced matrix element 
for the impulse term is
\begin{eqnarray}
\frac{1}{\sqrt{4\pi}}<\!p_f[^1S_0]\,||\,
{\cal T}^{\rm Imp}_{l_\pi}(q)
\,||\,p_i [^3P_0]\!>\; &=&\;
   \frac{-i}{\sqrt{(2\pi)^3 2\omega_q}}
\frac{g_A}{f_{\pi}}
\int \int\frac{d\vec{p}^{\,\,\prime}d\vec{p}}{4\pi}
  R_{^1S_0,p_f}(p') \nonumber \\
 &&  {\mbox{\hspace{-2cm}}}   \times \hat{p} 
\cdot (- \vec{q} + 
\frac{\omega_q}{m_N}\vec{p}^{\,\,\prime})
 \delta(\vec{p}^{\,\,\prime} - \vec{p} + \vec{q}/2)
  R_{^3P_0,p_i}(p)
\label{eq:TImppspace}
\end{eqnarray}
while the reduced matrix element
of the rescattering term is given by
\begin{eqnarray}
\frac{1}{\sqrt{4\pi}}<p_f[^1S_0]\,||\,
{\cal T}_{l_{\pi}=0}^{\rm Resc}(q)\,||\,p_i [^3P_0]>
 &=&  \frac{i}{\sqrt{(2\pi)^3 2\omega_q}}
\frac{2 g_A}{f_{\pi}}
\int \int\frac{d\vec{p}^{\,\,\prime}d\vec{p}}{4\pi}
  R_{^1S_0,p_f}(p') \nonumber \\
&\times& \frac{\kappa(k,q)}{(2\pi)^3}
\frac{\hat{p} \cdot \vec{k}}{ k^2 - m_{\pi}^2}
  R_{^3P_0,p_i}(p).
\label{eq:TRescpspace}
\end{eqnarray}
In the above,
$\vec{p}$ and $\vec{p}^{\,\,\prime}$ stand for
the relative momenta of the two protons
before and after the pion emission, respectively, and 
$k = (k_0,\vec{k}) =
(E_{\vec{p}}-E_{\vec{p}^{\,\,\prime} -
\vec{q}/2}, \vec{p} - \vec{p}^{\,\,\prime} + \vec{q}/2)$.
The radial functions for the $pp$ scattering states
that appear in Eqs.~(\ref{eq:TImppspace}),
(\ref{eq:TRescpspace}) are to be generated
with the use of realistic nucleon-nucleon interactions.
SLMK\cite{slmk97} used 
the Argonne V18 potential.\cite{wir95}\footnote{SLMK
checked that the use of, \eg, 
the Reid soft-core potential
gives essentially the same results.}
If we take the limit $ \vec{q} \rightarrow 0$ limit
and freeze $k_0$, the energy variable of the exchanged pion,
at the $fixed$ value $k_0=m_\pi/2$ corresponding
to the threshold pion production,
then we are back with the {\it TT kinematics} results.
This simplified treatment is equivalent 
to the {\it fixed kinematics approximation}
used in PM$^3$K.\cite{pmmmk96}

The upshot of Sato \etal's results\cite{slmk97}
is as follows.
The rescattering contribution
$<p_f[^1S_0]\,||\,
{\cal T}_{l_{\pi}=0}^{\rm Resc}(q)\,||\,p_i [^3P_0]>$
estimated in the full p-space calculation
has the same sign as the result 
of the {\it fixed kinematics approximation}
but, as far as their magnitudes are concerned,
\be
|\!<\!p_f[^1S_0]\,||\,
{\cal T}_{l_{\pi}=0}^{\rm Resc}(q)
\,||\,p_i [^3P_0]\!>\!|_{\rm full}
\approx
|\!<\!p_f[^1S_0]\,||\,
{\cal T}_{l_{\pi}=0}^{\rm Resc}(q)
\,||\,p_i [^3P_0]\!>\!|_{\rm fix.\,kin.}
\lab{eq:comparison}
\ee
In PM$^3$K, 
$<\!p_f[^1S_0]\,||\,
{\cal T}_{l_{\pi}=0}^{\rm Resc}(q)\,||\,p_i [^3P_0]\!>
\approx 
-<\!p_f[^1S_0]\,||\,
{\cal T}^{\rm Imp}_{l_\pi}(q)\,||\,p_i [^3P_0]\!>$,
leading to an almost perfect cancellation 
between the impulse and rescattering terms.
SLMK find instead
\be
<\!p_f[^1S_0]\,||\,
{\cal T}_{l_{\pi}=0}^{\rm Resc}(q)\,||\,p_i [^3P_0]\!>
\approx 
-3<\!p_f[^1S_0]\,||\,
{\cal T}^{\rm Imp}_{l_\pi}(q)\,||\,p_i [^3P_0]\!>
\lab{eq:resimp}
\ee
Therefore, although the rescattering and impulse terms 
interfere destructively in this case also,
the dominance of the rescattering term
leads to a much larger cross section than 
in the {\it fixed kinematics approximation}.
Yet, $\sigma_{\rm full}$ obtained in SLMK
is still significantly smaller
than the observed cross sections.

SLMK also examined to what extent
the uncertainties in the low-energy coefficients,
$c_1$, $c_2$ and $c_3$ affect their calculational results.
They found that the near-threshold 
$pp \rightarrow pp\pi^0$ cross sections
are sensitive to the value of $c_1$
and change significantly as $c_1$ varies
within the currently accepted error bars 
[see Eq.~(\ref{eq:lecoef})].
It was found, however, that even with the most
favorable choice of $c_1$ the calculated cross sections
are significantly smaller than the observed values.

To place the calculation of SLMK
in an appropriate context, 
the following three remarks are in order.

(i) ${\cal T}^{(+1)}$ in Eq.(\ref{eq:Tplus1})
represents only the tree-diagram contribution, Fig.1(b);
loop corrections to the $\nu=-1$ impulse term
generate transition operators of order $\nu=1$.
These additional contributions generate
an effective $\pi NN$ vertex form factor
for the impulse term, fig.1(a).
SLMK ignored these effects quoting a rough estimate
in PM$^3$K which indicates that 
the net effect of the loop corrections after renormalization
is less than $20 \%$ of the leading-order impulse term.
It is obviously desirable to reexamine this issue,
and we are currently carrying out a calculation
that includes all the $\nu=1$ contributions.\cite{sdmk}
Qualitatively speaking, the inclusion of this effect
is expected to reduce the contribution of
the impulse term, further enhancing the dominance
of the rescattering term.

(ii) The nuclear chiral counting scheme
employed above is in fact best applicable
when energy-momentum transfers
to a nucleus are small,
whereas the near-threshold
$pp \rightarrow pp\pi^0$ reaction involves
significant energy transfers $q_0\sim m_\pi$.
We therefore must exercise caution in applying
Weinberg's counting rule, Eq.(\ref{eq:Dcount}),
to this case.
In SLMK as well as in PM$^3$K, 
in the stage of constructing 
the transition operators,
the energy-momentum transfer 
due to the final pion is ignored
({\ie}, the external pion is taken to be soft,
$q_\mu\approx 0$)
in order to utilize Weinberg's original counting rule.
The physical value of $q_\mu$
is used only at the stage of calculating
the phase space integral.
The approximate nature of this approach is
particularly evident for the impulse term,
where the pion-emitting nucleon is
off-shell by $\sim m_\pi$.
This off-shell nucleon must interact
at least once with a second nucleon
before losing its off-shell character.
It is then sensible to treat
an impulse diagram accompanied
by subsequent one-pion exchange
as an irreducible diagram,
even though the original Weinberg classification
would treat it as a reducible diagram.
CFMV\cite{cfmv96} proposed
a modified chiral counting rule
that takes account of this feature.
In addition, these authors argued that
the $\Delta$ degree of freedom should be
taken into account explicitly in $\chi$PT
since the $N$-$\Delta$ mass difference $\sim 2m_\pi$
is small on the chiral scale $\Lambda$
(see also Ref.\cite{jm91a}).
All these are important issues
that deserve further investigations.
In particular, if one uses 
the CFMV's counting rule,\cite{cfmv96}
the expansion parameter is not any longer 
$m_\pi/M_{\mbox{\tiny N}}$
but $\sqrt{m_\pi/M_{\mbox{\tiny N}}}$.
This indicates a possible slow convergence of
chiral expansion.
(The insufficiency of the lowest-order rescattering diagram 
for describing $pp \rightarrow pp\pi^0$ was discussed 
in a different context in Ref.\cite{bla90}.)

(iii) This last remark is rather technical 
but it may contain an important message.
SLMK find that the rescattering transition matrix element 
is sensitive to uncomfortably high momentum components of
the nuclear ($pp$) wavefunction;
as a matter of fact, the relevant range of momentum
is not much smaller than the chiral scale $\Lambda$.
This disturbing feature is in fact shared 
also by other known applications of nuclear \CPT.
A satisfactory solution to this problem
probably requires the study of terms
with higher chiral orders than considered so far.

Despite all the caveats stated above,
it seems almost certain that,
in any reasonably realistic \CPT\ calculations,
the rescattering term dominates over the impulse term
and their signs are opposite to each other.
This implies that the heavy-meson
exchange contributions
considered in Ref.\cite{lr93}
cannot be invoked as a possible mechanism
to enhance $\sigma_{\rm full}$ to bring it closer
to the observed cross sections.
Since it is established that these
heavy-meson contributions
have the same sign as the impulse term,
the addition of these extra contributions
to the transition amplitude obtained in
the full p-space calculation would
result in a destructive interference.
Thus the heavy-meson contributions such as considered
in Ref.\cite{lr93} suppresses the cross section
instead of enhancing it.
Most recently, van Kolck, Miller and Riska\cite{kmr96}
considered yet another diagram involving
$\rho-\omega$ exchanges.
But, again, since this extra contribution has
the same sign as the impulse term,
one encounters the same difficulty as above.
Thus, the near-threshold
$pp \rightarrow pp\pi^0$ reaction
awaits further detailed investigations.

\section{Radiative muon capture on proton}

My third topic is concerned with the latest 
experiment on the radiative $\mu$-capture (RMC) on the proton: 
$\mu^-+p\ra n+\nu_\mu+\gam$.
It has long been a tremendous experimental challenge
to observe RMC on the proton, 
$\mu^-+p\rightarrow n+\nu_\mu+\gamma$,
because of its extremely small branching ratio.
Recently, an experiment at TRIUMF \cite{jonetal96} finally succeeded
in measuring $\Gamma_{\rm RMC}$,
the proton RMC rate.
To be more precise, the TRIUMF experiment
determined the partial capture rate 
$R(>60 {\rm MeV})$, corresponding to
emission of a photon with $E_\gamma>60$ MeV.
The matrix element of the hadronic charged weak current 
$h^\lambda=V^\lambda -  A^\lambda$
between a free proton and a free neutron
is given by 
\begin{eqnarray}
&\langle &\!\!
n(p_f) | V^\lambda - A^\lambda | p (p_i) 
\rangle  
\; =  
\nonumber \\
&{\bar {u}}& \!\!(p_f) \!\left[f_V(q^2)\gamma^\lambda
 + \frac{f_M(q^2)}{2m_N}
\sigma^{\lambda\mu}q_\mu
 + f_A(q^2)\gamma^\lambda\gamma_5
 + \frac{f_P(q^2)}{m_\pi}q^\lambda\gamma_5
\right] \!u(p_i),
\label{eq:formfactors}
\end{eqnarray}
where $q\equiv p_i-p_f$, and 
the absence of the second-class current is assumed.
Of the four form factors appearing 
in Eq.(\ref{eq:formfactors}),
$f_P$ is experimentally the least well known.
Although ordinary muon capture (OMC) on a proton, 
$\mu^-+p\rightarrow n+\nu_\mu$,
can in principle give information on $f_P$,
its sensitivity to $f_P$ is intrinsically suppressed.
This is due to the fact 
that the momentum transfer involved in OMC,
$q^2=-0.88 m_\mu^2$, 
is far from the pion-pole position $q^2=m_\pi^2$,
where the contribution of $f_P(q^2)$ 
becomes most important.
The radiative $\mu$-capture
provides a more sensitive probe of $\fP$ than OMC,
because the three-body final state in RMC
allows kinematical regions 
that are close to the pion pole.

To relate the measured RMC rate, $\GRMC$, to $\fP$,
the TRIUMF group\cite{jonetal96}
used a theoretical formula of Fearing.\cite{fea92}
Fearing's formula was derived essentially by invoking minimal substitution
to generate the transition matrix for RMC
from that for OMC.
Thus, after extracting the pion pole factor from $\fP$ 
to write $\fP(q^2)={\tilde {\fP}}/(q^2-m_\pi^2)$,
one replaces every $q$ in Eq.(\ref{eq:formfactors})
with $q-e{\cal A}$ 
(${\cal A}$ is the electromagnetic field)
except the $q$ appearing in the $q^2$ dependence 
in $\fV$, $\fA$ and $\fM$.
By treating $\GRMC^{\rm theor}$ obtained this way
as a function of ${\tilde {\fP}}$,
${\tilde {\fP}}$ is optimized to 
reproduce $\GRMC^{\rm exp}$.
(Speaking more precisely, 
the photon spectrum with a cut-off 
$E_\gamma\ge 60$ MeV was fitted.)
The result expressed in terms of 
$\gP\equiv \fP(q^2=-0.88m_\mu^2)$,
the value of $q^2$ relevant to OMC, is
$\gP=(10.0\pm0.9\pm0.3)\fA(0)$.
This value of $\gP$is $\sim1.5$ times the value
expected from PCAC.
This surprising result should be contrasted 
with the fact that $\gP$ measured in OMC
is consistent with the PCAC prediction
although the experimental uncertainties are large.

The result reported in Ref.\cite{jonetal96}
motivates us to ask: 
How reliable is $\GRMC^{\rm theor}$ 
used in deducing $\gP$ from $\GRMC^{\rm exp}$ ?
The theoretical framework of Fearing\cite{fea92}
used in Ref.\cite{jonetal96} is rather operational in character
and it is desirable to reassess its reliability.
For example, we know that the nucleon matrix element
of the charged weak current can contain covariants other 
than those that appear in Eq.(\ref{eq:formfactors}),
if the initial and/or final nucleons are off-shell.
This means that Eq.(\ref{eq:formfactors}) may be 
too restrictive for the description of RMC.
That is, for off-shell nucleons, 
there are many possible form factor expressions 
to which one could apply minimal substitution 
$q\rightarrow q-e{\cal A}$.
Then, even within the framework of minimal substitution,
there is ambiguity that cannot be lifted
with phenomenological approaches.\footnote{Dr. Cheon's talk at this meeting
addresses this type of problem.\cite{che97}}
In this context, even the statement that 
RMC is sensitive to the pseudoscalar formfactor $f_P$
should be taken with some caution.

Here again, a systematic calculation based on \CPT\  
is very useful
because (up to certain chiral orders) \CPT\ uniquely gives 
all the necessary vertices for electromagnetic coupling 
as well as for strong interactions.
Thus we can avoid applying 
a phenomenological minimal-coupling substitution
at the level of the transition amplitude. 
Furthermore, \CPT\ enables us to satisfy 
the gauge-invariance and chiral-symmetry requirements
in a transparent way.
Muon capture is a favorable case for applying \CPT\ 
since momentum transfers involved 
here do not exceed $m_\mu$, 
and $m_\mu$ is small compared to the chiral scale 
$\Lambda \sim $ 1 GeV, 
indicating the possibility of a reasonably 
rapid convergence of the chiral expansion.
In the case of OMC, Bernard \etal\cite{bkm94} 
and Fearing \etal\cite{feaetal97}
used heavy-baryon \CPT\ to evaluate $f_P$ 
with better accuracy than achieved 
in the PCAC approach. 
In the case of RMC, a \CPT\  calculation 
provides a natural extension 
of the classic work of Adler and Dothan\cite{ad66}
based on the low-energy theorems. 
For instance, the ${\cal O}(kq)$ terms that remained undetermined
within the framework of the low energy theorems
can be evaluated in \CPT.
We also note that,
in the case of threshold pion photo- and electroproduction,
chiral loop corrections lead to
a significant deviation of $E_{0+}$ 
from the low-energy theorem value.\cite{bkm95}

Recently, Thomas Meissner, Fred Myhrer and myself (MMK)
carried out a systematic \CPT\  calculation of 
$\GRMC^{\rm theor}$ 
to next-to-leading order.\cite{mmk97}
An extension of this treatment 
to sub-sub-leading order seemed to be a real challenge to us
but this has already been achieved by Ando and Min.\cite{am97}
Here I wish to describe briefly the work of the USC group.\cite{mmk97}
It is to be noted that
\CPT\  applied to this single-nucleon process
is free from the aforementioned extra complications
that afflict {\it nuclear} \CPT.

In the framework of the heavy-baryon \CPT,\cite{jm91}
MMK use the effective Lagrangian ${\cal L}_{\rm ch}$ 
in Eqs.(\ref{eq17a}), (\ref{eq17b}).
As explained earlier,
${\cal L}_{\rm ch}$ is written in the most general form 
involving pions and heavy nucleons
in external electromagnetic and weak fields 
consistent with chiral symmetry.
Limiting themselves to 
a next-to-leading chiral order (NLO) calculation,
MMK only keep terms 
with ${\bar{\nu}} = 0$ and ${\bar{\nu}} = 1$.
To this chiral order, 
only tree diagrams need to be considered,
and then ${\cal L}_{\pi N}^{(1)}$ simply represents
$1/M$ ``nucleon recoil'' corrections to 
the leading ``static'' part ${\cal L}_{\pi N}^{(0)}$. 

We consider all possible Feynman diagrams 
up to chiral order $\nu$ = 1 
which contribute to the process 
$\mu^- + p \to n + \nu + \gamma$.
The leptonic vertices in these Feynman diagrams are 
of course well known.
The hadronic vertices are obtained 
by expanding the \CPT\  Lagrangian ${\cal L}_{\rm ch}$
in terms of the elementary fields $B$, $\pi$, 
$\cal{V}$ and $\cal{A}$ and their derivatives. 
The evaluation of the transition amplitudes 
corresponding to these Feynman diagrams 
is straightforward,
and their explicit expressions can be found
in Ref.\cite{mmk97}.
MMK use the Coulomb gauge,
which assures $v \cdot \epsilon(\lambda) =0$.
The use of this relation
combined with a number of kinematical approximations
consistent with the accuracy 
of the $\nu=1$ ]CPT calculation
drastically simplifies the calculation
since many of the hadronic radiation diagrams
become ${\cal{O}}(1/M^2)$
and hence negligible.

Although this is certainly not a place 
for going into details,
several salient features of MMK's results
are worth emphasizing. 
The pion-pole diagrams originate from 
${\cal L}_\pi^{(0)}$ due to the coupling 
of the axial vector to the pion. 
In \CPT\  the pion-pole contributions,
which arise automatically 
from a well-defined chiral Lagrangian,
contain no ambiguity.
The fact that they need not be introduced by hand
constitutes a major advantage of the \CPT\  approach 
over the phenomenological approaches which have
been used in the earlier calculations.
MMK also report that some of the pion pole terms
that appear in \CPT expansion have no counterpart
in the phenomenological minimum substitution approach.
In this connection it is also worthwhile to mention 
that in \CPT calculations
the transition matrix for RMC
need not be directly related 
to the pseudoscalar coupling $g_P$ itself.
The Lagrangian ${\cal L}_{\rm ch}$ uniquely
determines $g_P$.\cite{bkm94,feaetal97}
as well as the RMC amplitude.
In this sense $g_P$ and $\Gamma_{\rm RMC}$
are of related to each other
but their relation is not as {\it trivial}
as indicated by the phenomenological treatment.
Furthermore, MMK observe that, 
with the use of the same Coulomb gauge,
the behavior of the contributions of certain diagrams
can differ between \CPT\ and 
the phenomenological treatment
(\eg, that in Ref.\cite{mw59}).

As far as numerical computations are concerned,
MMK\cite{mmk97} considered only the RMC 
from the $\mu p$ atomic state 
with the hyperfine states averaged over.
The spin-averaged total capture rate is given by
\begin{eqnarray}
\Gamma_{\rm RMC} \,  = \,  &&
\left ( 
\frac{eG}{\sqrt{2}} \right )^2
\vert \Phi (0) \vert^2
\,
\frac{1}{4}
\,
(2\pi)^4
\,
\int \frac{d^3 n}{(2\pi)^3}
\,
\int \frac{d^3 \nu}{(2\pi)^3}
\,
\int \frac{d^3 k}{(2\pi)^3}
\,
\frac{1}{2 \omega_k}
\nonumber \\ 
&& \times
\delta^{(4)} (n+\nu +k - p - \mu)
\sum_{\sigma{\sigma^\prime}s{s^\prime}\lambda} 
\vert M \vert^2 \; ,
\label{rmc1}
\end{eqnarray}
where the sum is over all spin 
and polarization orientations, and
$M = \sum_{i=1}^6 M_i$
with $M_i$ representing six distinct hadronic 
radiation amplitudes;
$\Phi(0)$ is the value of the $\mu p$ atomic wavefunction 
at the proton position.

The $\nu=1$ \CPT\ calculation of MMK\cite{mmk97}
gives the total capture rate 
$\Gamma_{\rm RMC}=0.075\,s^{-1}$,
of which $0.061\,s^{-1}$ 
comes from the leading-order 
${\cal{O}} ((1/M)^0)$ terms,
and $0.014\,s^{-1}$ from the ${\cal{O}} (1/M)$ terms  
due to ${\cal{L}}_{\pi N}^{(1)}$. 
If the contributions of the pion-pole diagrams
are dropped,
the resulting total capture rate would be 
$\Gamma_{\rm RMC}|_{{\rm no}\,\pi}=0.053\,s^{-1}$,
of which $0.043\,s^{-1}$ comes from 
the ${\cal{O}} ((1/M)^0)$ terms and 
$0.010\,s^{-1}$ from the ${\cal{O}} (1/M)$ terms.
Thus about 30 \% of the $\Gamma_{\rm RMC}$
comes from the pion-pole diagrams.
MMK's result for the total capture rate 
$\Gamma_{\rm RMC} = 0.075 s^{-1}$
is close to the value given in Ref.\cite{opa64}, 
$\Gamma_{\rm RMC}\,=\,0.069 s^{-1}$, 
and practically identical to 
$\Gamma_{\rm RMC}\,=\,0.076 s^{-1}$
reported in Ref.\cite{fea80}.
The ${\cal{O}} (1/M)$ recoil corrections 
are found to account for about $20\%$ 
of the leading order 
${\cal{O}} ((1/M)^0)$ contribution,
which indicates a reasonable convergence 
of the chiral expansion.  
By contrast, the size of the $1/M$ corrections 
is noticeably larger in the approach of Ref.\cite{fea80}. 

MMK\cite{mmk97} did not consider capture
from the singlet and triplet hyperfine states separately,
or capture from the $p\mu p$ molecular state.
Their results therefore cannot be directly compared
with the experimental data \cite{jonetal96}
I also repeat that MMK included
only up to the next-to-leading chiral order (NLO) 
contributions.
To perform next-to-next-to-leading order (NNLO) calculations,
one must include the $\bar{\nu} =2$ chiral Lagrangian,
${\cal{L}}_{\pi}^{(2)}$ and ${\cal{L}}_{\pi N}^{(2)}$,
and also loop corrections arising from ${\cal{L}}_{\pi}^{(0)}$ 
and ${\cal{L}}_{\pi N}^{(0)}$. 
Since chiral expansion for muon capture is
characterized by the expansion parameter $m_\mu/M$,
we expect a reasonably rapid convergence.
Indeed, in the case of OMC, 
where the $\nu=2$ calculation is much less involved,
explicit evaluations\cite{bkm94,feaetal97} show 
that the NNLO contributions amount only to a few percents.
This feature is likely to persist for RMC as well, 
but it is reassuring to check it explicitly.
It should also be mentioned 
that the formalism of Bernard \etal\cite{bkm95} 
used by MMKused does not contain 
the explicit $\Delta$ degree of freedom
in contrast to the approaches of Ref.\cite{jm91}.
The inclusion of the $\Delta$ particle
may turn out to be more important
than the NNNO extension.

Ando and Min's calculation for RMC\cite{am97}
not only includes the NNLO contributions
but also project out the hyperfine states
for both atomic and molecular capture.
As far as the spin-averaged $\Gamma_{\rm RMC}$
is concerned, the result of Ando and Min is
very close to that of MMK, 
indicating that chiral expansion for RMC indeed
converges rapidly.
Ando and Min conclude 
that $\Gamma_{\rm RMC}^{\rm exp}$
reported in Ref.\cite{jonetal96} is extremely difficult
to understand from the \CPT\.point of view.
In this situation it seems important to me 
to examine the contribution of the $\Delta$ particle,
which is still missing in the existing \CPT\ calculations
for RMC.
It is also to be mentioned that
Fearing and his collaborators are now undertaking 
a most systematic \CPT\ evaluation of RMC.\cite{}

\section*{Acknowledgments}
It is my great pleasure to participate 
in this Workshop organized to celebrate 
Dr. Mannque Rho's 60th birthday.
I wish to take this opportunity 
to express my deep gratitude to Mannque
for his warm friendship 
over more than a quarter of a century.
This work is supported in part
by the National Science Foundation,
Grant No. PHYS-9602000.

\end{document}